GIBBONFINDR: An R package for the detection and classification of acoustic signals

Dena J. Clink[1*] and Holger Klinck[1]

[1] *Center for Conservation Bioacoustics, Cornell Lab of Ornithology, Cornell University, Ithaca, NY*

* Corresponding author

Email: djc426@cornell.edu

## Abstract:
The recent improvements in recording technology, data storage and battery life have led to an increased interest in the use of passive acoustic monitoring for a variety of research questions. One of the main obstacles in implementing wide scale acoustic monitoring programs in terrestrial environments is the lack of user-friendly, open source programs for processing large sound archives. Here we describe the new, open-source R package GIBBONFINDR which has functions for detection, classification and visualization of acoustic signals using a variety of readily available machine learning algorithms in the R programming environment. We provide a case study showing how GIBBONFINDR functions can be used in a workflow to detect and classify Bornean gibbon (*Hylobates muelleri*) calls in long-term acoustic data sets recorded in Danum Valley Conservation Area, Sabah, Malaysia. Machine learning is currently one of the most rapidly growing fields-- with applications across many disciplines-- and our goal is to make commonly used signal processing techniques and machine learning algorithms readily available for ecologists who are interested in incorporating bioacoustics techniques into their research.

## Introduction:
Researchers worldwide are becoming increasingly interested in passive acoustic monitoring (PAM), or the use of autonomous recording units to monitor vocal animals and their habitats. The increase in availability of low-cost recording units (e.g. Koch et al., 2016; Hill et al., 2018; Sethi et al., 2018), along with advances in data storage capabilities makes the use of PAM an attractive option for monitoring vocal species in inaccessible areas where the animals are hard to monitor visually (such as dense rainforests) or when the animals exhibit cryptic behavior [4]. Even in cases where other methods such as visual surveys or trapping are feasible, PAM may be superior as it may be able to detect animals continuously for extended periods of time, at a greater range than visual methods, can operate under any light conditions, and is more amenable to automated data collection than visual or trapping techniques [5].

One of the most widely recognized benefits of using acoustic monitoring, apart from the potential to reduce the amount of time needed for human observers, is that there is a permanent record of the acoustic data [7]. In addition, the use of archived acoustic data allows for multiple observers at different times to analyze and validate classifications, as opposed to point-counts where the data are generally taken by one or a few observers, and data are ephemeral. In many cases, analysis of recordings taken by autonomous recorders can be more effective than using trained human observers in the field. For example, a comparison of the ability of automated detection based on recordings taken using autonomous recorders and the ability of human observers to detect European nightjars (*Caprimulgus europaeus*) showed that automated detection methods detected nightjars during 19 of 22 survey periods, while surveyors detected nightjars on only six of these occasions [7]. A recent analysis of 21 bird studies that compared detections by human observers and autonomous recorders found that for 15 of the studies autonomous recorders performed equally well or better than humans [8]. Despite the rapidly expanding advances in PAM technology, the use of PAM is limited by a lack of widely applicable analytical methods and limited availability of open-source audio processing tools, particularly for the tropics where soundscapes are very complex [6].

## Detection and classification of signal(s) of interest

One of the most pressing issues facing the wide-scale adoption of PAM is the need for reliable recognition of the signal(s) of interest from long-term recordings [9,10]. Particularly with the advances in data storage capabilities and deployment of arrays of recorders collecting data continuously, the amount of time necessary for hand-browsing or listening to recordings for signals of interest is prohibitive, and is not consistent with conservation goals that require rapid assessment. The development of different automated detection approaches for terrestrial animals is an active area of research [11–15]. Given the diversity of signal types and acoustic environments there is no single detection algorithm that performs well across all signal types and recording environments.

There are no standardized or widely available programs that can be used for the automated processing of the raw sound files that are produced from handheld or autonomous digital recordings [16], although there are many programs for the manual inspection of spectrograms and isolation of signals of interest such as Raven [17], Praat [18], along with signal processing functionalities in popular programming languages such as R [19] and Python [20]. Signals of interest can either be detected manually using visual or aural techniques, or using automated methods. Spectrogram cross-correlation is a commonly used method that can be used to identify signals of interest from long-term recordings [12], although the use of spectrogram cross-correlation requires relatively high signal-to-noise ratio and relatively low variability within the call type of interest. Generally, processing acoustic data consists of four steps: identification of the signal of interest (detection), feature extraction, classification and validation (Figure 1).

Depending on the type of signal detection method used, along with the overall goal of the PAM program, there will most likely be a need to subsequently classify signals of interest into particular categories (e.g. sexes, individuals). If automated detection of signals is used, generally the first step will be binary classification of the signal(s) of interest and background noise. Then, subsequent classification into categories of interest can be done if needed. With the increasing popularity of machine learning techniques, ecologists have begun to use a variety of classification algorithms. Some of the more common algorithms used for human speech recognition such as hidden markov models [21], support vector machines (SVMs: [13,22]), and neural networks [23] have been applied to the automated detection of animal calls. When doing automated detection of animal calls, the number and type of training data must be taken into consideration, with the aim of reducing the number of false-positives (where the system falsely classifies a signal as the signal of interest), or false negatives (e.g. missed opportunities), where the system fails to detect the signal of interest.

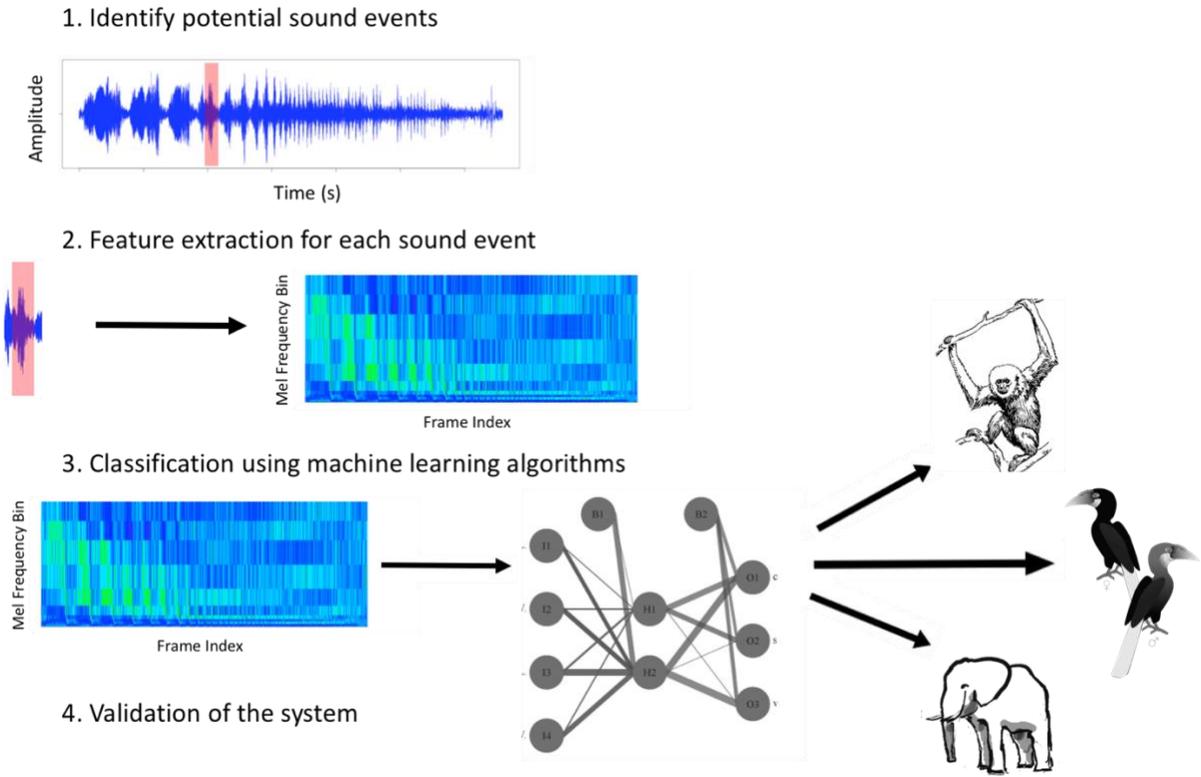

**Figure 1. Schematic of automated detection workflow.**

## Feature extraction

Direct analysis and synthesis of the acoustic signal using the raw sound is generally not feasible, given the complexity and amount of data [24].The sampling rate, bit size and duration of a recording will determine the size of the sound file, but regardless of the size of the sound file, digital samples of acoustic signals generally have high amounts of redundant information. The process of feature extraction is used as a general data reduction technique, wherein values are estimated from the original dataset; chosen values are meant to be informative and non-redundant [25]. The most common method of feature extraction in primate studies is to estimate temporal and spectral features from the spectrogram (e.g. chimpanzees, Marler and Hobbett, 1975; macaques, Gouzoules and Gouzoules, 1990; orangutans, Lameira and Wich, 2008; and gibbons, Terleph et al., 2015; Clink et al., 2017). Feature extraction from the spectrogram is often subjective, and generally very time consuming.

Another commonly used feature extraction is the use of Mel-Frequency Cepstral coefficients (MFCCs), which are commonly used in human speech recognition [31–33]. MFCCs are based on the "Mel-scale", which more closely aligns with pitch perception in terrestrial vertebrates [34]. Animals perceive changes in frequency below 1000 Hz linearly, but that is not the case at frequencies above 1000 hz, which means that the linear scale tends to overemphasize high-frequency components of vocalizations [35]. MFCCs have been used to classify marmoset [37], blue monkey [38], gibbon [39], orangutan [40] and chimpanzee [41] vocalizations. MFCCs are less suitable for testing hypotheses related to call production and evolution, as differences in the coefficients are difficult to interpret in a biologically meaningful way [38], but are effective for classification problems. A representative spectrogram and MFCC coefficient plot is shown in Figure 2.

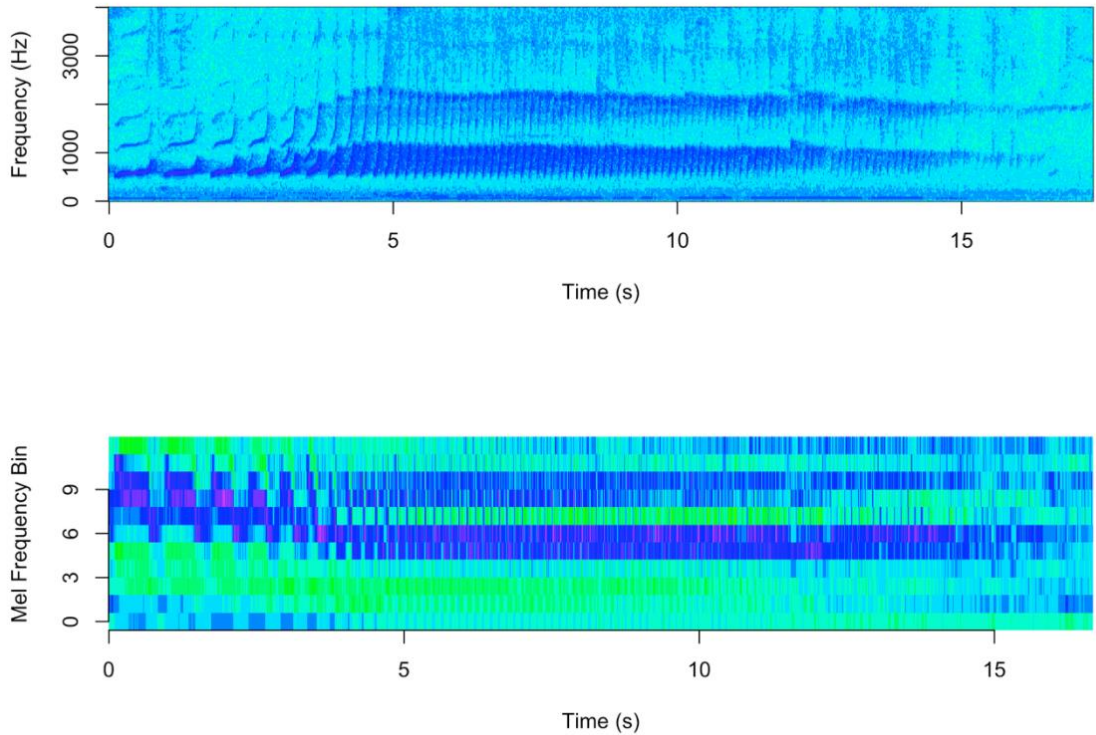

**Figure 2. A representative spectrogram of a Bornean gibbon female call (top) and Mel-frequency cepstral coefficients (MFCCs) for the same call (bottom).** Spectrograms are commonly used for visualization of acoustic signals, and MFCCs are commonly used features for detection and classification problems [36].

## Validation

A crucial component for any classifier is validation. Validation of the classifier should be done with a different dataset than that which was used to train the models, to avoid the possibility of overfitting [22,42]. Some commonly used metrics include precision (the proportion of detections that are true detections) and recall (the proportion of actual calls that are successfully detected [42]). Often, metrics are converted to hourly rates, such as the rate of false positives per hour, which can help guide decisions about type of detector and classifier to use. In addition, when doing automated detection and classification it is common to use a threshold (such as the probability assigned to a classification by a machine learning algorithm) to make decisions about rejecting or accepting a detection [42]. Varying these thresholds will result in changes to false-positive and proportion of missed calls, and these can be plotted with receiver operating curves (ROC; [43]) or detection error tradeoff curves (DET; [44]). A hypothetical confusion matrix for classification of gibbon calls, which shows how different categorization metrics (e.g. False positives) are calculated, is shown in Table 1.

**Table 1. Confusion matrix for classification of gibbon calls.**

|                       | *Gibbon*       | *Not Gibbon*   |
|----------------------:|----------------|----------------|
| *Test says Gibbon*     | True Positive  | False Positive |
| *Test says Not Gibbon* | False Negative | True Negative  |

# Package summary

Machine learning, a fast growing field in computer science, is a form of artificial intelligence that "learns" from training data to perform particular tasks, such as classification of detected signals [45]. Artificial neural networks [38], Gaussian mixture models [22] and Support Vector Machine [13,22] -- some of the more common algorithms used for human speech recognition [24,32] – can be used for the automated detection of terrestrial animal signals from long-term recordings. Our goal with GIBBONFINDR is to provide an open-source, step-by-step approach with thoroughly annotated code for the automated detection and classification of acoustic signals using machine learning. This package is aimed towards ecologists who are interested in incorporating acoustic approaches into their research program. This package complements existing R packages for acoustic analysis such as TUNER [46], SEEWAVE [47], WARBLER [48] and MONITOR [12], and contributes functionalities for automated feature extraction using Mel-frequency cepstral coefficients, detection using machine learning algorithms, and visualization of detections.

## Data preparation

Training machine learning algorithms requires human- labeled training data. Labeling training data can be achieved in one of two ways. First, the user can manually browse the spectrograms and isolate signals of interest using programs with a GUI interface such as Praat [18] or Raven Pro [17]. Alternatively, the user can use audio segmentation on longer sound files to isolate potential sound events (see audio segmentation section below), and provide class labels for each of the sound events. We provide datasets that consist of labelled sound files of nine distinct gibbon females ("gibbon.females") that were obtained via hand-browsing of spectrograms of high-quality focal recordings taken in Danum Valley in 2015 [49]. In addition, we provide sound files of gibbon females of lower signal-to-noise ratio, gibbon males, leaf monkeys and noise ("multi.class.list") that were obtained using the "detectSNR" function from a long-term recording in Danum Valley Conservation Area, Sabah, Malaysia using a Swift autonomous recorder [3] and classified by a human observer (DJC). We also provide five 15-minute recordings taken in February 2018 using Swift autonomous recorders [3] set at a ~750 m spacing. All datasets are available on GitHub (https://github.com/DenaJGibbon/gibbonR-package).

## Visualization and processing of training data

In GIBBONFINDR we provide functions for the feature extraction and visualization of multiple labeled sound events. The function "calcMFCC" calls on the package TUNER [46] to calculate MFCCs. Briefly, the calculation of MFCCs is as follows. First, a Fast Fourier Transform is calculated for each time window, and then the frequency axis is converted to the Mel-scale by using a series of logarithmically spaced filters, and the energy from the frequency filters is input to a discrete cosine transform, which outputs cepstral coefficients [36]. The "calcMFCC" function will calculate the MFCCs for each sound event in two different ways. First, the function will calculate MFCCs for each time window (similar to how spectrograms are calculated), which will result in a MFCC vector for each sound event that is of variable length, assuming that sound events are of different duration. Second, the function will divide each sound event into a user-specified number of time windows, and calculate MFCCs for each time window, resulting in a standardized number of MFCCs for each sound event [38,39]. The function "plotSoundevents", which relies on the package SIGNAL [52] will read in files from a specified input directory and print them to a plot for visual inspection (see Figure 3 for a representative output).

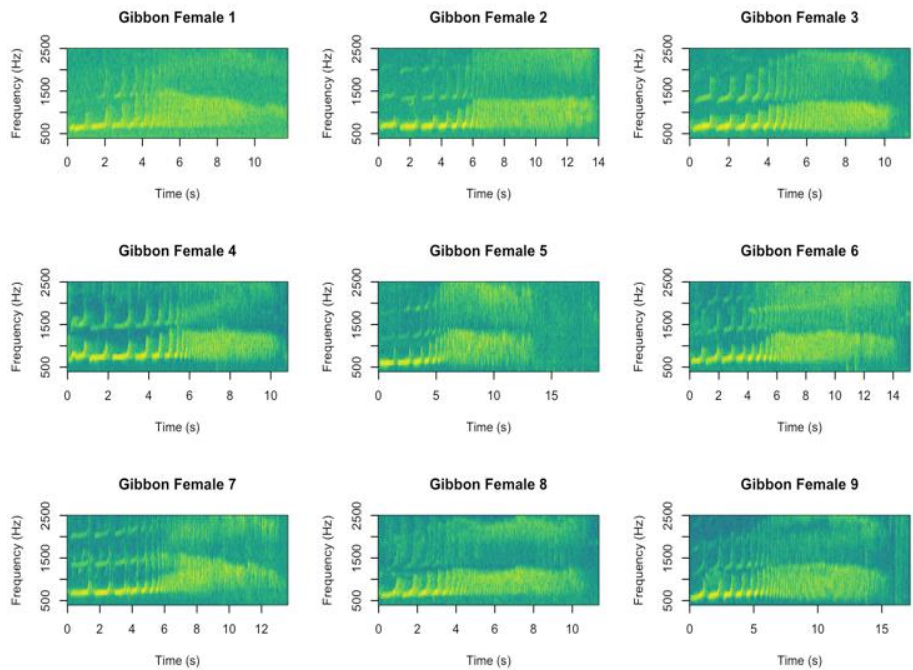

**Figure 3. Representative output of the function "plotSoundevents" which allows user to plot multiple spectrograms to visualize acoustic data.** Spectrograms show individual calls from nine distinct female gibbons taken during focal recordings [49] from the 'gibbon.females' dataset provided with the package. Spectrograms were made with a 1600-point Hanning window and 0% overlap.

## Training and testing machine learning algorithms

The R programming environment has many readily available machine learning algorithms, and we provide functions that call on these packages to train and test different machine learning algorithms using user-labeled training data (see Table 2 for list and description of functions). The machine learning functions rely on the packages E1071 [53], CARET [54] and MCLUST [55] to train support vector machines (SVM [56]), neutral networks (NN [23]) and gaussian mixture models (GMM [57,58]), respectively. These functions use the default settings of the package developers, which means they are not optimized, and we urge users to modify the settings so that classification is optimized for their particular research problem. The user can define the percentage of data that is used for testing and training, and each function returns a confusion matrix and a percent correct classification. We also provide a function that relies on GGPLOT2 [59] to plot a biplot of the results of either principal component analysis or linear discriminant function analysis (see Figure 4 for a representative biplot).

**Table 2. List of functions in the GIBBONFINDR package**

| *Function* | *Description* |
| --- | --- |
| *Data processing* | |
| detectSNR | A detector that identifies potential sound events based on signal-to-noise ratio |
| detectSVM | A detector that identifies potential sound events using a trained SVM |
| calcMFCC | Calculate Mel-frequency cepstral coefficients for sound events |
| *Machine learning* | |
| trainGMM | Use a Gaussian mixture model to classify labelled sound events |
| trainLDA | Use linear discriminant function analysis to classify labelled sound events |
| trainNNET | Use a neural network to classify labelled sound events |
| trainSVM | Use a SVM to classify labelled sound events |
| classifyGibbonR | Classifies unlabeled sound events using trained machine learning algorithms |
| *Data visualization* | |
| biplotGibbonR | Create a biplot based on principle component analysis or linear discriminant function analysis |
| calldensityPlot | Create a call density plot based on the number of detections at specified recording locations |
| plotSoundevents | Create spectrograms of sound events in a directory |
| soundEventSpec | Create spectrogram of sound file with sound events detected by gibbonFindR outlined |
| *Automated detection* | |
| gibbonFindR | Function to detect and classify signal(s) of interest from long-term recording(s) |

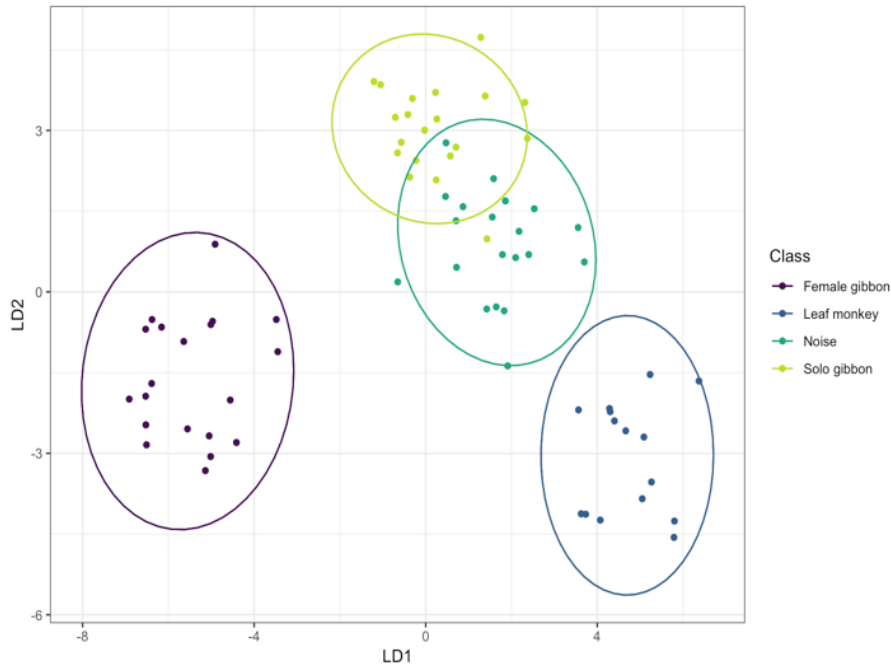

**Figure 4. Representative biplot of the first two linear discriminant functions created with the function "biplotGibbonR".** Each sound event was isolated from a long-term recording using the 'detectSNR' function, and manually labeled by a human observer (provided with the package in the 'multi.class.list' dataset). MFCCs were calculated for each sound event using the 'calcMFCC' function.

## Classification of unlabeled sound events

The function "classifyGibbonR" allows the user to specify an input directory with .wav files of potential sound events and classify the events using either SVM, NN or GMMs. This function requires the user to input training data.

## Detection

Detectors are commonly used to isolate potential sound events of interest from background noise [60–62]. In GIBBONFINDR we provide functions to identify potential sound events based on band-limited energy summation [42] and SVMs [60,62]. For the band-limited energy method, we first create a spectrogram of the audio recording with non-overlapping time windows and filter to the frequency range of the signal of interest (in the case of Bornean gibbons 0.4- 1.8. kHz). We then calculate the sum of energy in each of the time windows, and based on user-specified duration and quantile values identify signals that are above a certain threshold and duration. This function does not require that user to input training data. The second function relies on a trained SVM and does require the user to input training data. For this detector, MFCCs are calculated for the entire audio recording, and for each time window the trained SVM classifies the window as "gibbon" or "noise", and then based on user-specific probability threshold and duration, identifies series of time windows that constitute a sound event.

## Putting it all together: automated detection

Automated detection of signals generally follows four main steps: 1) identification of potential sound events using audio segmentation; 2) data reduction and feature extraction of sound events; 3) classification of sound events using trained machine learning algorithms; and 4) validation of the system. When training the system it is important to use data that will not be used in the subsequent testing phase, as this may artificially inflate accuracy estimates [22]. Training the system generally relies on the use of observer annotated data along with subsequent training of the classification algorithm based on signal classes defined by the observer in the training data. There are no clear-cut guidelines in terms of the choice of algorithm, and it is common for authors to test the effectiveness of multiple algorithms for classification of signals of interest.

GIBBONFINDR provides a function for the automated detection and classification of acoustic signals ("gibbonFindR"). This function requires the user to input training data, and to specify a target class or signal type (e.g. female gibbon) and it allows the user to specify whether they want to use SVM, NN or GMM algorithms for classification. The function follows the following steps. First, it trains the specified machine learning algorithm with user input training data. Then, it detects potential sound events using band-limited energy summation or SVMs. The function then calculates MFCCs for each sound event and classifies the sound events using the trained machine learning algorithm. The function can output either cut .wav files into a specified directory, or a table with MFCC features and a class label, and it also returns a table with the times of the sound events. The user can visualize individual sound events using the "plotSoundevents" function or create a spectrogram with the sound events outlined using the "soundEventSpec" function (Figure 5).

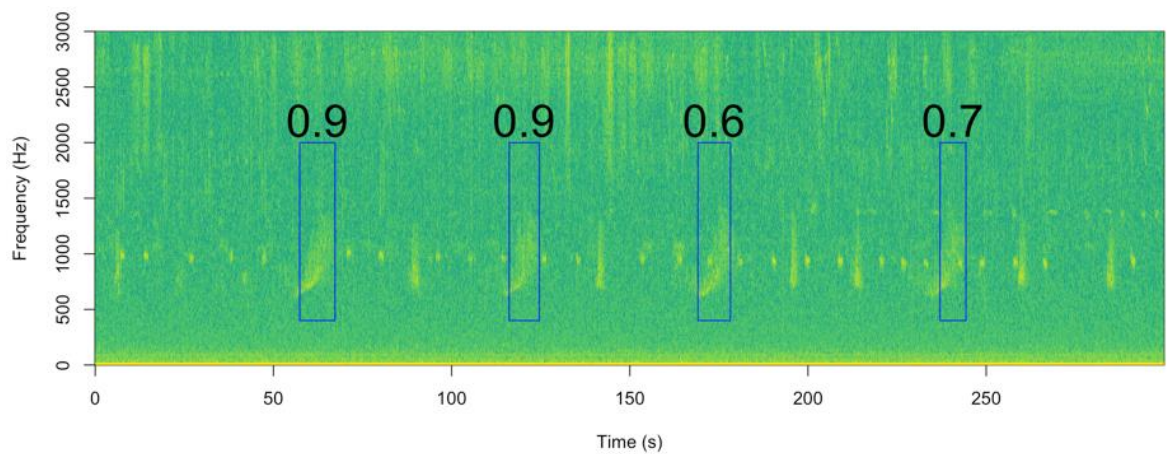

**Figure 5. Spectrogram of recording taken at Danum Valley Conservation Area.** The blue boxes show the detections of the female contribution to the duet with "gibbonFindR", and the numbers above the boxes are the probabilities that the signal belongs to the class assigned by SVM.

## Call density plot

PAM can provide spatial information about the presence or absence of vocal animals, if autonomous recorders are set on an array that encompasses sufficient area that calls are detected on some recorders but not others [10]. We provide a function that calculates a call density surface using inverse distance weighted interpolation from the package GSTAT [63] and returns a plot of the call density surface (Figure 6). Call detections can be input from the automated detector provided in this package, or a table with call detections based on user annotations. This allows for users to investigate spatial patterns in detections of their target signal or species over an array of autonomous recorders.

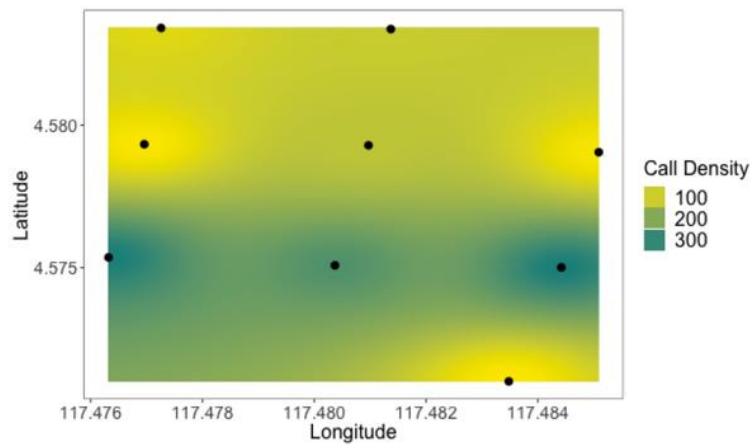

**Figure 6. Bornean gibbon female call density surface across the study area at Danum Valley Conservation Area, Sabah Malaysia.** The inverse distance weighted interpolation is based on total number of calls at each recording location, and the black points denote recorder locations.

# A case study on gibbons in Danum Valley Conservation Area
## Case Study: Methods
### Data collection

We deployed ten Swift autonomous recording units [3] spaced on a 750-m grid encompassing an area of approximately 3 km$^2$ in the Danum Valley Conservation Area, Sabah, Malaysia (4°57'53.00"N, 117°48'18.38"E) from February 13- April 21, 2018. Recorders were attached to trees at a height of approximately 2m, and recorded continuously over a 24-hour period. Recordings were saved as consecutive two-hour Waveform Audio File Format (WAV) with a size of approximately 230 MB. We recorded using a gain of 40 dB and a sampling rate of 16 kHz, giving a Nyquist frequency of 8 kHz, which is well above the range of Bornean gibbon calls (0.4 to 1.8 kHz). We limited our analysis to recordings taken between 06:00 and 12:00, as gibbons tend to limit their calling to the early morning hours [64,65], which resulted in a total of over 4500 hours of recordings for the automated detection. See Clink et al. [65] for a detailed description of study design.

We focused our analysis the female contribution to the duet, known as the great call, for two reasons. First, the structure of the great call is highly stereotyped, individually distinct

[26,27], of longer duration than other types of gibbon vocalizations, and the males tend to be silent during the female great call, which facilitates better automated detection. Second, most density estimation techniques focus on the great call, as females rarely sing if they are not in a mated pair, whereas males will solo whether they are in a mated pair or if they are drifters [66], which means automated detection of the female call will be more relevant for density estimation using passive acoustic monitoring.

**Training data**

It is necessary to validate automated detection systems using different training and test data sets [22], so we used recordings taken during the same time period but from two different recording units located within our larger array. We used a total of 112 hours of training data (from 14 mornings) and used the band-limited energy detector (settings described below) to identify potential sounds of interest in the gibbon frequency range, which resulted in 4126 unique sound events. The subsequent sound events were then annotated by a single observer (DJC) in the following categories: argus, barking deer, bird, cicada, gibbon, gibbon female great call, gibbon male solo, hornbill, human, leaf monkey, noise, and rain. We included the generic class "gibbon" for all gibbon vocalizations that were not a complete female call or male solo. To augment our training data, we also included 209 gibbon female calls that were collected during focal recordings during previous field seasons at our site [49]. For simplicity of training the machine learning algorithms we converted our training data into two categories "female gibbon" or "not female gibbon", and we subsequently trained binary classifiers, although the classifiers can also deal with multi-class classification problems.

**Detection**

We were interested to compare the performance of two different detectors (band-limited energy and SVM). For the band-limited energy detector we first converted each two-hour recording to a spectrogram (made with a 1600-point (100 ms) Hamming window (3 dB bandwidth = 13 Hz), with 0% overlap, and a 2048-point DFT) using the package 'seewave' [47]. We then filtered the spectrogram to the frequency range of interest (between 0.4 and 1.5 kHz), and for each non-overlapping time window we calculated the sum of the energy across frequency bins, which resulted in a single value for each time window. We used the 'quantile' function in base R to calculate the threshold value for signal versus noise. We ran early experiments using different qualities thresholds and found that using the 50% quantile gave the best separation for our signals of interest. We then considered any events which lasted for a duration of 6-sec or longer to be detections.

For the SVM detector, we first trained a SVM with the R package 'e1071' [53] using labeled sound events from the 'multi.class.list' dataset outlined above. We calculated MFCCs for each sound event using the following settings: window size 0.25-sec., 12 MFCCs per window calculated between 0.4 and 1.5 kHz. Then we assigned a category (gibbon or not) and a probability for each time window using the trained SVM and considered any events that were assigned to the 'gibbon' category that were over 6-sec duration to be detections.

## Classification

We were interested to test the performance of secondary classifiers (either SVM, GMM or NN) for classifying our detected sound events. To train the classifiers we used the training data set outlined above, and calculated MFCCs for each of the labeled sound events using a slightly different method than that used for the detector. We averaged MFCCS over time windows as the duration of the sound events is variable, and the machine learning algorithms require a feature vector of equal length for each sound event. First we divided each sound event into 9 evenly spaced time windows (with the actual length of each window varying based on the duration of the sound event), and calculated 12 MFCCs along with the delta coefficients for each time window using the package 'tuneR' [46]. We appended the duration of the event onto the MFCC vector, resulting in a vector for each sound event of length 177. We then used E1071 [53] to train a SVM, CARET [54] to train a NN and MCLUST [55] to train a GMM. Each of the algorithms assigned each sound event to a class ('female gibbon' or 'not gibbon') and associated probability.

## Validation

To validate our detector and classifiers one observer (DJC) manually annotated 230 randomly chosen hours of recordings (which accounted for 5% of total recording time) taken from different recorders and times across our study site using spectrograms created in RavenPro [17], identifying the begin and end time of any female gibbon vocalization and labelling calls as high quality (wherein the full structure of the call was visible in the spectrogram and there were no overlapping birds or background noise) or low quality (wherein the call was visible in the spectrogram but the full structure was not, or there was overlapping with another calling animal). We validated our detectors using two different approaches. First, for each detector and classifier we calculated the recall (the proportion of gibbon calls in the recording that were detected) and the hourly false positive rate (the number of sound events that were incorrectly classified as gibbons divided by the total number of recording hours).

We were also interested to see how the performance of our classifiers varied when we changed the probability threshold, so we calculated ROC curves. ROC curves show the trade-off between the rate of false-positives and false-negatives, where the True Positive Rate (TPR) is equal to:

$$\frac{True\ Positive}{True\ Positive + False\ Negative}$$

And the False Positive Rate (FPR) is equal to:

$$\frac{False\ Positive}{False\ Positive + True\ Negative}$$

# Case Study: Results
### Validation

We found that there was a tradeoff between the hourly false positive rate and recall using different detectors, call quality thresholds and machine learning algorithms (Figure 6), with both detectors (band-limited energy and SVM) having higher recall but also higher false positive rate than when a secondary classifier is applied. Secondary classification using Gaussian mixture models tends to have the lowest false positive rate but also the lowest recall, whereas secondary classification with SVM has higher recall with a slightly higher false positive rate (Figure 7). We also found that varying the probability threshold of the three classifiers lead to a tradeoff between the false positive rate and true positive rate (Figure 8).

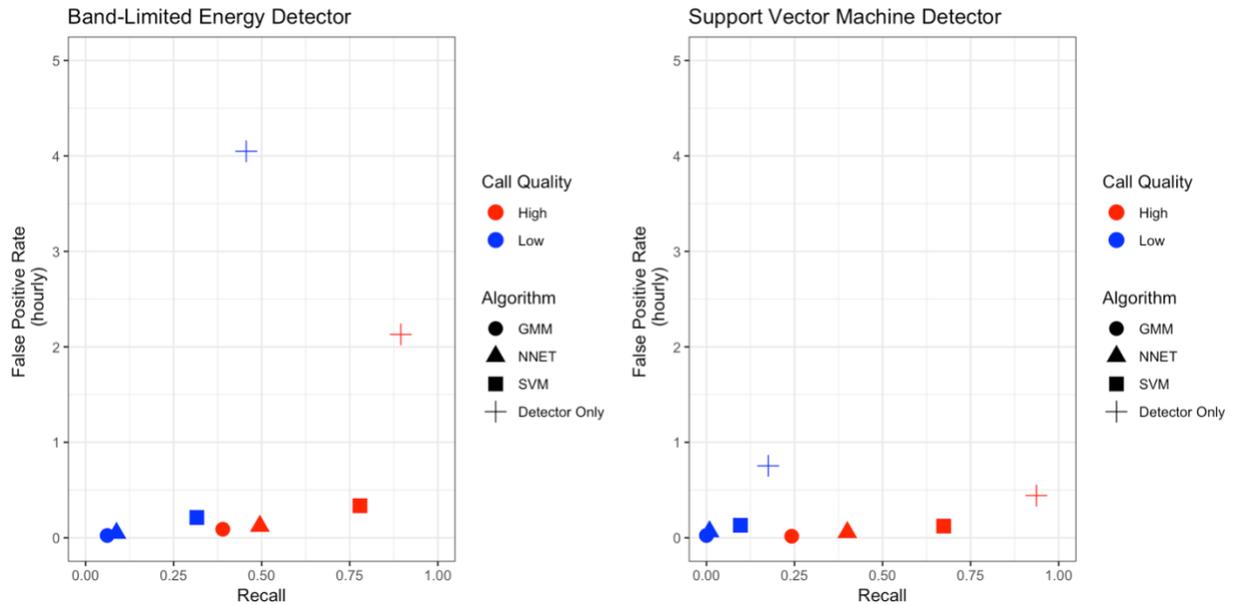

**Figure 7. Recall and False Positive Rate (hourly) for two different detector types (Band-limited energy and Support Vector Machine), three different classifiers (Gaussian Mixture Models, Neural Networks and Support Vector Machines) for high- and low-quality gibbon female calls.**

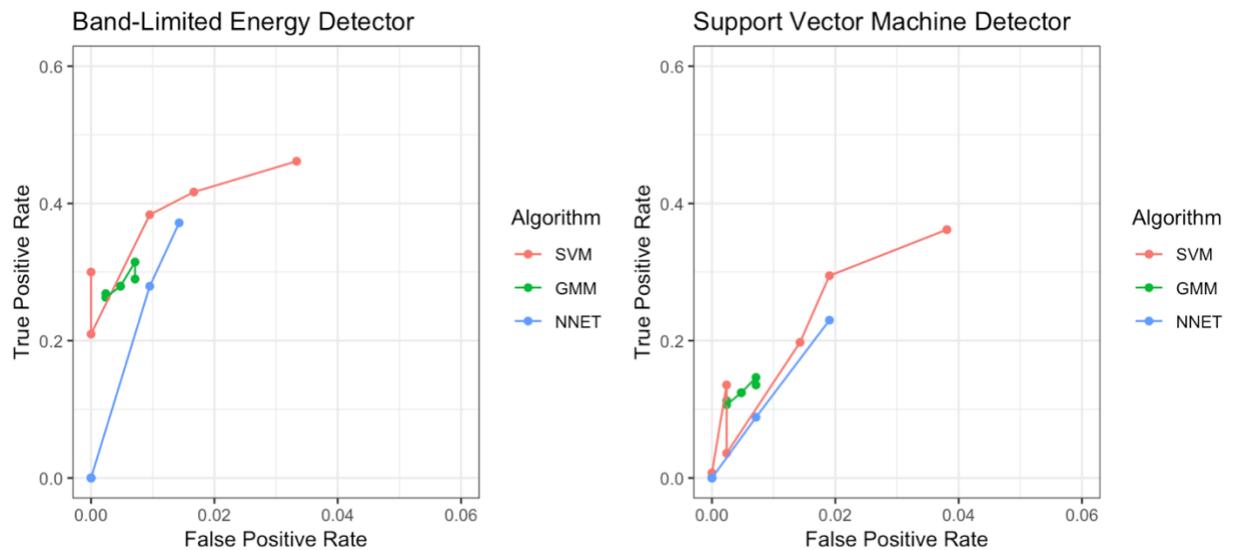

**Figure 8. Representative ROC curve showing the tradeoff between False Positive Rate and True Positive Rate at different probability thresholds (0%, 50%, 75%, 85%, 95% and 99%).**

## Case Study Discussion

We show that using open-source R packages, a detector and classifier can be developed with an acceptable performance that exceeds that of previously published automated detectors for primate calls [22]. We found that using only the detectors led to the highest recall, but unsurprisingly the highest false positive rate, whereas the use of Gaussian mixture models led to the lowest false positive rate, but also had the lowest recall. The tolerable amount of false positives, or the minimum tolerable recall of the system, will depend heavily on the research question. For example, when doing occupancy modeling it may be important that no calls are missed, so a higher recall would be desirable. But, for studies that focus more on the behavioral ecology of the calling animals, it may be important for the detector to identify calls with a low amount of false positives, but less important if the detector misses many low signal-to-noise calls.

Gibbon female calls are well-suited for automated detection and classification as they are highly stereotyped, and gibbons tend to call often, and during a particular calling bout the emit multiple calls, which allows for ample training data. Although gibbon female calls have been shown to be individually distinct [27,39], the differences between individuals were not sufficient to preclude detection and classification using our system. The generalizability of our methods to other systems will depend on a variety of conditions, in particular the signal-to-noise ratio of the call(s) of interest relative to background noise, the amount of stereotypy in the calls of interest, and the amount of training data that can be obtained to train the system.

## Conclusions

Our goal in the creation of this package was to highlight how the open-source R-programming environment can be used for the processing and visualization of acoustic data collected using autonomous recorders that are often programmed to record continuously for long periods of time. Even the most sophisticated machine learning algorithms are never 100% accurate or precise, and will return false positives or negatives [13,15,22], which is also the case with human-observers but is rarely quantified statistically [22]. Modifying the parameters of the machine learning algorithms- in particular programming the algorithms to return probabilities of class membership- can help the user determine the acceptable amount of false positives or negatives for their particular research question. The algorithms included in this R package are far from optimized; they are implemented using the default values set by the algorithm developers. We have thoroughly annotated the source code and invite more advanced R users to refer to the vignettes of the relevant packages and modify as they see fit for their particular research problem.

## Acknowledgements

We gratefully acknowledge Allison R. Lau for helpful input on earlier versions of "GIBBONFINDR". We also thank the makers of the packages "CARET", "E1017", "MCLUST", "SEEWAVE", "SIGNAL", and "TUNER", on which this package relies extensively.

*Authors' contributions*
DJC conceived the ideas and designed the methodology, and DJC and HK led the writing of the manuscript. Both authors contributed critically to the drafts and gave final approval for publication.

*Works cited*
All code for R package and data needed to recreate analyses are on available on Github (https://github.com/DenaJGibbon/ gibbonR-package).